\documentclass[a4paper]{article}

\usepackage{INTERSPEECH2021}
\usepackage{amsmath,graphicx}
\usepackage{amssymb,amsfonts,url}
\usepackage{algorithmic}
\usepackage{textcomp}
\usepackage[rgb]{xcolor}
\usepackage{float}
\usepackage{placeins}
\usepackage{psfrag}
\newcommand{\stz}{\rule{0mm}{2.4ex}}

\usepackage{color,soul,cite}
\usepackage{multirow}
\usepackage{makecell}
\usepackage{xcolor,colortbl,hhline}
\usepackage[super]{nth}
\usepackage{subcaption}
\usepackage{setspace}
\usepackage{dblfloatfix}

\title{Deep Noise Suppression With Non-Intrusive PESQNet Supervision\\
Enabling the Use of Real Training Data}
\name{Ziyi Xu, Maximilian Strake, Tim Fingscheidt}
\address{Institute for Communications Technology, Technische Universit{\"a}t Braunschweig, Germany}
\email{$\left \{ \text{ziyi.xu, m.strake, t.fingscheidt} \right \}$@tu-bs.de}
\begin{document}
\ninept
\maketitle
\begin{abstract}
Data-driven speech enhancement employing deep neural networks (DNNs) can provide state-of-the-art performance even in the presence of non-stationary noise. During the training process, most of the speech enhancement neural networks are trained in a fully supervised way with losses requiring noisy speech to be synthesized by clean speech and additive noise. However, in a real implementation, only the noisy speech mixture is available, which leads to the question, how such data could be advantageously employed in training. In this work, we propose an end-to-end non-intrusive {\tt PESQNet} DNN which estimates perceptual evaluation of speech quality (PESQ) scores, allowing a reference-free loss for real data. As a further novelty, we combine the {\tt PESQNet} loss with denoising and dereverberation loss terms, and train a complex mask-based fully convolutional recurrent neural network (FCRN) in a ``weakly" supervised way, each training cycle employing some synthetic data, some real data, and again synthetic data to keep the {\tt PESQNet} up-to-date. In a subjective listening test, our proposed framework outperforms the Interspeech 2021 Deep Noise Suppression (DNS) Challenge baseline overall by $0.09$ MOS points and in particular by $0.45$ background noise MOS points.
\end{abstract}
\section{Introduction}\label{sec:intro}
Speech enhancement aims at improving intelligibility and perceived quality of a speech signal degraded by disturbances, which can include both additive noise and reverberation. The classical solution for single-channel denoising is to estimate a time-frequency (TF) domain mask for spectral weighting \cite{Ephraim1984,Scalart1996,Cohen2005a,Gerkmann2008b,fodor2012mmse,samy_SNR}.

Data-driven approaches (e.g., deep learning) have facilitated the direct estimation of TF-domain masks or the desired clean speech spectrum, which provide state-of-the-art performance, even in the presence of non-stationary noise \cite{fingscheidt2006data,Fingscheidt2008,wang2014training,weninger2014discriminatively,wang2015deep,park2017fully,zhao2018convolutionalrecurrent,tan2019complex,strake2019separated,strake2020fully}. Convolutional neural networks (CNNs) are well-suited to preserve the harmonic structures of the speech spectrum in denoising tasks \cite{park2017fully,zhao2018convolutionalrecurrent,tan2019complex,strake2019separated,strake2020fully}. Integrating long short-term memory (LSTM) into CNNs can further improve the denoising performance as shown in recent studies \cite{zhao2018convolutionalrecurrent,tan2019complex,strake2019separated,strake2020fully,braun2020data}. Strake et al.\ \cite{strake2020fully} proposed a fully convolutional recurrent neural network (FCRN), combining CNNs with convolutional LSTM (ConvLSTM) layers, which inherits the weight-sharing property of the CNN, thereby improving performance over standard LSTM.

The Deep Noise Suppression (DNS) Challenge for Interspeech 2020 (DNS1) \cite{reddy2020interspeech}, ICASSP 2021 (DNS2) \cite{reddy2020icassp}, and Interspeech 2021 (DNS3) \cite{reddy2021interspeech} addressed a challenging task for joint dereverberation and denoising. Strake et al.\ \cite{strake2020DNS} achieved a second rank in the DNS1 non-realtime track by training an actually realtime-capable FCRN for complex spectral masking-based denoising and dereverberation, which we build upon. All of the abovementioned works apply supervised training using synthetic data, where clean speech and noise are prepared separately and are mixed by addition, while reverberation (if considered) is applied by convolution with simulated or recorded room impulse responses (RIRs). However, {\it in a real-world implementation, only the noisy speech mixture is available, leading to the question, how such real data could be advantageously employed in training.} A semi-supervised training for speech enhancement deep neural networks (DNNs) is proposed in \cite{pascual2017segan,pandey2018adversarial,meng2018cycle}, which utilize both synthetic and real data. All of these works employ generative adversarial networks (GANs), using a generator to separate the target signal from the mixture, while a discriminator \mbox{distinguishes} between the true clean speech signal and the enhanced signals. Thus, real data can be used for training the generator, while synthetic data, where clean speech must be available, is used for training the discriminator. Furthermore, well-performing speech enhancement GANs usually still use an additional discriminative loss term such as an L1 loss to stabilize generator training \cite{pascual2017segan,pandey2018adversarial}, which can only be employed for synthetic data. {\it Beyond GAN-type losses, to the best of our knowledge, it has not yet been shown how real data can be used to train a speech enhancement DNN.}

To allow real data for non-GAN losses, a possible way is to train a network which predicts absolute category rating (ACR) listener scores, as, e.g., perceptual evaluation of speech quality (PESQ) \cite{ITUT_pesq_wb_corri} does it. However, the original PESQ is a non-differentiable function, which cannot be used as an optimization criterion for gradient-based learning. Zhang et al.\ \cite{zhang2018training} implemented an approximated gradient descent for the original PESQ, while {M}art{\'\i}n-{D}o{\~n}as et al.\ \cite{martin2018deep} proposed an approximated PESQ formulation, which is differentiable. Zhao et al.\ \cite{zhao2019perceptual} proposed a supervised psychoacoustic loss, which outperforms the approximated PESQ \cite{martin2018deep}. Fu et al.\ \cite{fu2019learning} trained an end-to-end DNN approximating the PESQ function (so-called {\tt Quality-Net}) without having to know any computation details of the original PESQ formulation. The trained {\tt Quality-Net} is used to provide a differentiable PESQ loss for the training of speech enhancement DNNs. However, as the original PESQ \cite{ITUT_pesq_wb_corri}, both \cite{martin2018deep,fu2019learning} require a clean reference signal, which makes them only suitable for a fully supervised training approach. Furthermore, the {\tt Quality-Net} is fixed during the enhancement model training, which strongly limits the performance, since the {\tt Quality-Net} can be fooled by the speech enhancement model (estimated PESQ scores increase while true PESQ scores decrease) after training for several minibatches, as reported by the authors of \cite{fu2019learning}.

In our work, we propose an end-to-end {\it non-intrusive} {\tt PESQNet} modeling the PESQ function, and subsequently we employ it to train a speech enhancement DNN. {\it Our proposed {\tt PESQNet} can estimate the PESQ score of the enhanced signals without knowing the corresponding clean speech} (just like human raters in ACR listening tests), thus, it can serve to provide a reference-free perceptual loss for training with real data. Furthermore, we solved the problems addressed in \cite{fu2019learning} by training the FCRN and the {\tt PESQNet} alternatingly, similar to alternating training schemes used in adversarial trainings \cite{pascual2017segan,pandey2018adversarial,meng2018cycle}. We combine the loss provided by the {\tt PESQNet} and the successful FCRN loss from \cite{strake2020DNS}, thereby training a complex mask-based FCRN in a ``weakly" supervised way, employing both synthetic and real data in training.

The rest of the paper is structured as follows: Section 2 introduces the notations, the speech enhancement system, and our novel weakly supervised training scheme with {\tt PESQNet}. The experimental setup as well as the results and discussion are presented in Section 3. We conclude the work in Section 4.
\vspace*{-1mm}
\section{Speech Enhancement With PESQNet}
\subsection{Deep Noise Suppression (DNS) by an FCRN}
We assume the microphone mixture $y(n)$ to be composed of the clean speech signal $s(n)$ reverberated by the room impulse response (RIR) $h(n)$, and disturbed by additive noise $d(n)$ as
\vspace*{-1mm}
\begin{equation} \label{micro_mixture}
\vspace*{-1mm}
y(n)=s(n)*h(n)+d(n)= s^\text{rev}(n)+d(n),
\end{equation}
with $s^\text{rev}(n)$ and $n$ being the reverberated clean speech component and the discrete-time sample index, respectively, and $*$ denoting a convolution operation. Since we perform a spectrum enhancement, we transfer all the signals to the discrete Fourier transform (DFT) domain (see Fig.\,\ref{fig:system}):
\vspace*{-1mm}
\begin{equation} \label{micro_fft}
Y_\ell(k)=S^\text{rev}_\ell(k)+D_\ell(k),
\vspace*{-1mm}
\end{equation}
with frame index $\ell$ and frequency bin index $k\!\in\!\mathcal{K}\!=\!\left \{0,1,\ldots,K\!-\!1\right \}$, and $K$ being the DFT size. Following \cite{strake2020DNS}, we use an FCRN structure delivering a magnitude-bounded complex mask $M_\ell\left(k \right )\in\mathbb{C}$, with $\left|M_\ell\left(k \right )\right|\in\left [ 0,1 \right ]$ to obtain the enhanced speech spectrum:
\vspace*{-1mm}
\begin{equation} \label{clean_speech_est}
\hat{S}_\ell\left (k \right )=Y_\ell(k)\cdot M_\ell\left(k \right ).
\vspace*{-1mm}
\end{equation}
Finally, the enhanced speech spectrum $\hat{S}_\ell\left (k \right )$ is subject to an inverse DFT (IDFT), followed by overlap add (OLA) to reconstruct the estimated signal $\hat{s}(n)$.

We adopt the fully-supervised loss function operating in the complex spectrum domain proposed in \cite{strake2020DNS}, which consists of two loss terms. The utterance-wise joint loss term aims at joint dereverberation and denoising by utilizing the clean speech spectrum $S_\ell(k)$ as target, and is defined as
\vspace*{-1mm}
\begin{equation} \label{joint}
J^\text{joint}_u\!=\!\frac{1}{L\!\cdot\!K}\!\sum_{\ell\in\mathcal{L}_u}\sum_{k\in\mathcal{K}}\!\bigl|\hat{S}_\ell(k)\!-\!S_\ell(k)\bigr|^2,
\vspace*{-1mm}
\end{equation}
with $\mathcal{L}_u$ being the set of frames belonging to the utterance indexed with $u$, and $L\!=\!\bigl|\!\mathcal{L}_u\!\bigr|$ being the number of frames. A second loss term only focuses on denoising by employing the reverberated clean speech spectrum $S^\text{rev}_\ell(k)$ as target in
\vspace*{-1mm}
\begin{equation} \label{noise}
J^\text{noise}_u\!=\!\frac{1}{L\!\cdot\!K}\!\sum_{\ell\in\mathcal{L}_u}\sum_{k\in\mathcal{K}}\!\bigl|\hat{S}_\ell(k)\!-\!S^\text{rev}_\ell(k)\bigr|^2\!.
\vspace*{-2mm}
\end{equation}
Finally, the two loss terms \eqref{joint} and \eqref{noise} are weighted by $\beta$, and combined in the {\it final loss for synthetic data} shown in Fig.\,\ref{fig:system}:
\vspace*{-1mm}
\begin{equation} \label{MT}
J^\text{synth}_u\!=\beta\cdot J^\text{joint}_u+(1-\beta)\cdot J^\text{noise}_u,
\vspace*{-2mm}
\end{equation}
with $\beta=0.9$, as used in \cite{strake2020DNS}.
\begin{table*}[!t]
	\scriptsize
	\centering
	\caption{\textbf{\textit{Instrumental quality results}} on the DNS1 Challenge \textbf{\textit{preliminary test set}} $\mathcal{D}^\mathrm{synth, dev}_\mathrm{DNS1}$, evaluated separately on \textbf{\textit{synthetic data}} without and with reverberation. DNSMOS is adopted from \cite{reddy2020dnsmos}. Best results are in {\bf bold} font, and the second best are \underline{underlined}.}
	\vspace*{-2mm}
	\setlength\tabcolsep{4pt}
	\begin{tabular}{ccc c c c c c c c}
		\hline
		&\multirow{2}{*}{Methods} & \multicolumn{4}{c}{\stz{Without reverb}}& \multicolumn{4}{c}{\stz{With reverb}}\\ \cmidrule(r){3-6} \cmidrule(r){7-10} 
		&    &  PESQ & DNSMOS  &  \stz{STOI}  &   $\Delta\text{SNR}_\text{seg}$[dB]   &    PESQ & DNSMOS   & \stz{STOI} &  SRMR \\ \hhline{----------} 
		\multirow{3}{*}{\rotatebox{90}{REF\ }}& \multicolumn{1}{l}{Noisy}             &  \stz{2.21} &  3.15  &  0.91   &  -     &  1.57  &  2.73  &0.56    &-   \\ \hhline{~~~~~~~~~~}
		& \multicolumn{1}{l}{\stz{DNS3 Baseline \cite{braun2020data}, trained with synthetic data only}}              &     3.15   &  3.64    &    0.94   &    {6.30} & 1.68   &  {\bf 3.18}   & \underline{0.62}   & 6.33\\ \hhline{~~~~~~~~~~}
		& \multicolumn{1}{l}{\stz{FCRN, with loss \eqref{MT} \cite{strake2020DNS}, trained with synthetic data only}}  &      {\bf 3.37}   &  {3.82}   &    {\bf 0.96}   &    8.35   & {\bf 1.95} &  3.08    &  {\bf 0.63}  & {7.25}\\ \hhline{----------}
		\multirow{3}{*}{\rotatebox{90}{New\ }} &\multicolumn{1}{l}{\stz{FCRN/PESQNet, losses \eqref{total} and \eqref{PESQNet}, $\langle0\!-\!2\!-\!50\rangle$ (synth.\ data only)}}   &    \underline{3.35}  &  {\bf 3.88}    &   \underline{0.95}  &    \underline{8.40} & \underline{1.92} &  3.11   &  {0.61}  & {\bf 7.41}\\ \hhline{~~~~~~~~~~}
		&\multicolumn{1}{l}{\stz{FCRN/PESQNet, losses \eqref{total} and \eqref{PESQNet}, $\langle1\!-\!1\!-\!50\rangle$ , $\alpha\!=\!0.8$}}   &    3.22  &   \underline{3.85}  &   \underline{0.95}   &  8.30   &  1.87 &   3.11  & 0.61   & \underline{7.00}\\ \hhline{~~~~~~~~~~}
		&\multicolumn{1}{l}{\stz{FCRN/PESQNet, losses \eqref{total} and \eqref{PESQNet}, $\langle1\!-\!1\!-\!50\rangle$, $\alpha\!=\!0.9$}}   &    3.29  &  {\bf 3.88}   &    \underline{0.95}  &    {\bf 8.52} & \underline{1.92}  &  \underline{3.14}   &  \underline{0.62}  & {6.85}\\ \hhline{----------}
	\end{tabular}
	\label{DNS1_dev}
	\vspace*{-4mm}
\end{table*}
\vspace*{-1mm}
\subsection{Novel PESQNet}
\label{sec:2.1}
Furthermore, we propose an end-to-end non-intrusive {\tt PESQNet} to estimate the PESQ score of an enhanced speech utterance in the DFT domain. This network shall allow the use of real data to train the FCRN by providing learning guidance without requiring a clean reference signal. Perceptual speech quality estimation delivers a single value (label) for an entire utterance, just as speech emotion recognition (SER) does. Furthermore, models used for SER are always non-intrusive, which is suitable for our reference-free PESQ estimation. So we adapted the SER model proposed in \cite{Meyer2021}, which shows state-of-the-art performance in emotion recognition, to construct our {\tt PESQNet}. The only difference to the original SER model in \cite{Meyer2021} is that we change the input to the network from log-mel spectrum to the amplitude spectrum, and the last output layer is changed from four output nodes with softmax activation function to a single output node with a gate function, which limits the estimated PESQ between $1.04$ and $4.64$, as it is done in \cite{ITUT_pesq_wb_corri}. All other settings can be taken from \cite{Meyer2021}. Note that any other SER model could be used alternatively.

\begin{figure}[t!]
	\ninept
	\vspace*{-2mm}
	\psfrag{A}[cc][cc]{RIR}
	\psfrag{B}[cc][cr]{$S_\ell(k)$}
	\psfrag{C}[cc][cc]{$D_\ell(k)$}
	\psfrag{D}[cc][cc]{NORM}
	\psfrag{E}[cc][cc]{FCRN}
	\psfrag{F}[cc][cc]{PESQ ITU-T P.862}
	\psfrag{G}[cc][cc]{$\hat{S}_{\ell}\left (k \right )$}
	\psfrag{H}[cc][cc]{PESQNet loss \eqref{PESQNet}}
	\psfrag{I}[cc][cc]{$\text{PESQ}(u)$}
	\psfrag{J}[cc][cc]{Total loss \eqref{total}}
	\psfrag{K}[cc][cc]{PESQNet}
	\psfrag{L}[cc][cc]{$\widehat{\text{PESQ}}(u)$}
	\psfrag{M}[cc][cc]{Synth.\ loss \eqref{MT}}
	\psfrag{N}[cc][cc]{Real loss \eqref{real}}
	\psfrag{O}[cc][cc]{$J^\text{total}_u$}
	\psfrag{P}[cc][cc]{$J^\text{synth}_u$}
	\psfrag{Q}[cc][cc]{$J^\text{real}_u$}
	\psfrag{R}[cc][cc]{$S^\text{rev}_\ell(k)$}
	\psfrag{S}[cc][cc]{$Y_\ell(k)$}
	\psfrag{T}[cc][cc]{$J^\text{PESQNet}_u$}
	\centering
	\centerline{\includegraphics[width=0.48\textwidth]{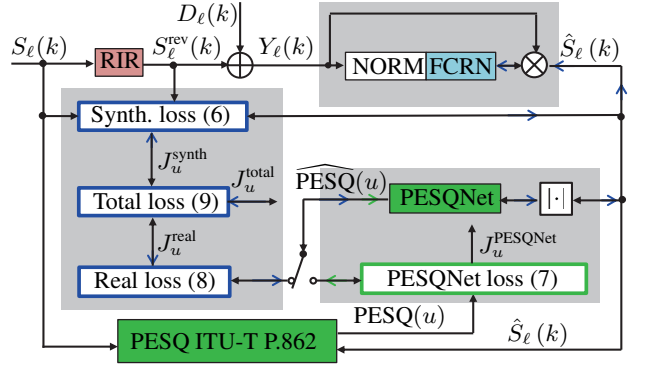}}
	\vspace*{-1mm}
	\caption{{\bf PESQNet and FCRN training setup.} The switch is used to alternately control the training of the FCRN (left switch position) or the PESQNet (right switch position).}
	\label{fig:system}
	\vspace*{-6mm}
\end{figure}

The output of the {\tt PESQNet} should be as close as possible to its ground truth measured by the original PESQ function \cite{ITUT_pesq_wb_corri}. Thus, the loss function for {\tt PESQNet} training is defined as
\vspace*{-1mm}
\begin{equation} \label{PESQNet}
J^\text{PESQNet}_u\!=\left(\widehat{\text{PESQ}}(u)-\text{PESQ}(u)\right)^2,
\vspace*{-1mm}
\end{equation}
with $\text{PESQ}(u)$ being the true PESQ score of the utterance indexed with $u$, measured by ITU-T P.\ 862 \cite{ITUT_pesq_wb_corri}, see Fig.\,\ref{fig:system}. Note that $J^\text{PESQNet}_u$ \eqref{PESQNet} can only be employed with synthetic data.
\vspace*{-1mm}
\subsection{Novel Weakly Supervised Training with PESQNet}
\label{sec:2.2}
\vspace*{-1mm}
The pre-trained {\tt PESQNet} is employed at the output of the FCRN estimating the PESQ scores of the enhanced speech, which should be maximized. Therefore, we can define a loss function
\vspace*{-2mm}
\begin{equation} \label{real}
J^\text{real}_u\!=\left(\widehat{\text{PESQ}}(u)-4.64\right)^2
\end{equation}
for utterance $u$, which is minimized during FCRN training as shown in Fig.\,\ref{fig:system}. {\it Please note that} $J^\text{real}_u$ \eqref{real} {\it is a speech enhancement loss, which for the first time, allows the use of real data beyond GAN-type losses.} Furthermore, for synthetic data only, we can explicitly consider joint dereverberation and denoising by combining the synthetic loss \eqref{MT} with the novel reference-free psychoacoustic loss \eqref{real}. Thus, our proposed final loss function for utterance $u$ in the weakly supervised training employing both real and synthetic data is
\vspace*{-1mm}
\begin{equation} \label{total}
J^\text{total}_u= 
\begin{cases}
J^\text{real}_u,\ &\text{for real data}\ \mathcal{D}^\text{real}\\
\alpha\cdot J^\text{synth}_u\!+\!(1\!-\!\alpha)\!\cdot\!J^\text{real}_u,\ &\text{for synthetic data}\ \mathcal{D}^\text{synth},
\end{cases}
\vspace*{-1mm}
\end{equation}
with $\alpha\in\left [ 0,1 \right ]$ being the weighting factor.

To prevent a fixed {\tt PESQNet} from being fooled by the updated FCRN (estimated PESQ scores increase while true PESQ scores decrease \cite{fu2019learning}), we use an alternating training protocol for adapting the FCRN with \eqref{total} and the {\tt PESQNet} with \eqref{PESQNet}. The FCRN and {\tt PESQNet} training is illustrated in Fig.\,\ref{fig:system}.  A zero-mean and unit-variance normalization based on the statistics collected on the training set is represented by the NORM box in Fig.\,\ref{fig:system}. The FCRN and the {\tt PESQNet} are trained alternately, controlled by the switch in left and right position, respectively. The left grey box containing the total loss $J^\text{total}_u$ \eqref{total} is used for FCRN training with a fixed {\tt PESQNet}. The lower-right grey box shows the {\tt PESQNet} training adapting it to the current fixed FCRN, employing $J^\text{PESQNet}_u$ \eqref{PESQNet}. The blue and green arrows indicate the gradient flow back-propagated for FCRN and {\tt PESQNet} training, respectively.
\vspace*{-1mm}
\section{Experimental Evaluation}
\subsection{Datasets, Methods, Training, and Metrics}
\label{sec:3.1}
\vspace*{-1mm}
In this work, signals have a sampling rate of $16\, \text{kHz}$ and we apply a periodic Hann window with frame length of $384$ with $50\%$ overlap, followed by an FFT with $K=512$. We directly adopted the FCRN from \cite[Fig.\,2]{strake2020DNS} as our DNS model, with exactly the same settings, as we do not focus on network topologies here. Note that the number of input and output frequency bins is $K_{\rm in}=257+3=260$. The last 3 frequency bins are redundant and only used for compatibility with the two maxpooling and upsampling operations, and are dropped for subsequent processing. With these settings (24 ms frame size and 12 ms frame shift), we measured a real-time factor of $r=0.95$ on an {\tt Intel Core i5 quad core} machine (3.4 GHz clock), meeting both real-time and the algorithmic delay restrictions ($24\,\text{ms}\! +\!12\,\text{ms}\!=\!36\,\text{ms}$ $\leq$ 40 ms) of the Interspeech 2021 DNS Challenge \cite{reddy2021interspeech}.

First, we pre-train the FCRN model using the WSJ0 speech corpus \cite{Garofalo2007} (dubbed $\mathcal{D}^\text{synth}_\text{WSJ0}$) and loss \eqref{MT}. Secondly, the {\tt PESQNet} is pre-trained on the same dataset with the enhanced speech spectrum $\hat{S}_\ell(k)$ generated by the pre-trained and fixed FCRN, using loss \eqref{PESQNet}. After the pre-trainings, we propose a two-stage fine-tuning with subsets of the data from DNS3 (denoted as $\mathcal{D}^\text{synth}_\text{DNS3}$). In a first stage, we fine-tune the pre-trained FCRN with the synthetic loss \eqref{MT} on the $\mathcal{D}^\text{synth}_\text{DNS3}$ dataset and subsequently, we fine-tune the {\tt PESQNet} on the same dataset with the enhanced speech spectrum generated by the fixed fine-tuned FCRN, again with loss \eqref{PESQNet}.

The second-stage (final) fine-tuning is used to advantageously employ real data for a weakly supervised training. Thus, besides the synthetic dataset $\mathcal{D}^\text{synth}_\text{DNS3}$, we also constructed a real dataset $\mathcal{D}^\text{real}_\text{DNS1+2}$ comprising the real recordings from both preliminary and blind test datasets from DNS1 and DNS2, in total $4$ hours of training material and $0.4$ hours of validation material. Only for the model which is used to process the final blind test data $\mathcal{D}^\mathrm{test}_\mathrm{DNS3}$ for a subjective listening test, we employ a real dataset with additional real data from the DNS3 {\it development} set (not blind {\it test} set of course), denoted as $\mathcal{D}^\text{real}_\text{DNS1+2+3}$. A cycle of our novel training protocol (second stage) is as follows: We first train the FCRN with a fixed {\tt PESQNet}, employing one minibatch (3 utterances) of {\it real} data followed by one minibatch of {\it synthetic} data using loss \eqref{total}. The minibatch with synthetic data not only explicitly considers denoising and dereverberation, but also prevents the FCRN from generating additional artifacts in the enhanced speech spectrum. Then, we fix the FCRN, and train the {\tt PESQNet} with $50$ consecutive minibatches of {\it synthetic} data using \eqref{PESQNet}, with the enhanced speech obtained from the current FCRN as shown in Fig.\,\ref{fig:system}. This training protocol is dubbed $\langle1\!-\!1\!-\!50\rangle$. We further explore performance when replacing the {\it real} minibatch data by an additional {\it synthetic} minibatch, which is dubbed $\langle0\!-\!2\!-\!50\rangle$. The weighting factor $\alpha$ in $J^\text{total}_u$ \eqref{total} is set to $0.9$ for all trainings, only for the novel training protocol $\langle1\!-\!1\!-\!50\rangle$ we perform an extra experiment with $\alpha=0.8$. During second-stage fine-tuning, for either the proposed protocol $\langle1\!-\!1\!-\!50\rangle$ or for the purely synthetic reference protocol $\langle0\!-\!2\!-\!50\rangle$, the FCRN model and the corresponding {\tt PESQNet} are saved after every $39$ protocol runs. Among all the saved FCRN models, we select the one which provides the lowest loss \eqref{total} on the respective validation sets of real or synthetic data. In our case, measurements on {\it both} real and synthetic validation sets lead to the same finally chosen model.

The pre-training dataset $\mathcal{D}^\text{synth}_\text{WSJ0}$ is exactly the same as used in \cite{strake2020DNS}, which is synthesized with clean speech from the WSJ0 corpus \cite{Garofalo2007} and noise from DEMAND \cite{thiemann2013diverse} and QUT \cite{dean2010qut}. We simulate SNR conditions of $0$, $5$, and $10$ dB according to ITU-T P.56 \cite{ITU56}. No reverberation effects are considered in preparation of the pre-training dataset. 

The fine-tuning dataset $\mathcal{D}^\text{synth}_\text{DNS3}$ contains $100$ hours of training material and $10$ hours of validation material with a uniformly sampled SNR between $0$ and $40$ dB. We reverberate $50\%$ of the files in $\mathcal{D}^\text{synth}_\text{DNS3}$ by convolving the clean speech component with simulated RIRs following \cite{strake2020DNS}.

We use the preliminary synthetic test set from DNS1 \cite{reddy2020interspeech} (dubbed $\mathcal{D}^\text{synth, dev}_\text{DNS1}$), which contains synthetic noisy speech mixtures with and without reverberation for instrumental performance measurement. For the noisy mixtures without reverberation, we measured the segmental SNR improvement $\Delta\text{SNR}_\text{seg}$ \cite{loizou2013speech}. To evaluate the dereverberation effects, we measured the speech-to-reverberation modulation energy ratio (SRMR) \cite{falk2010non} under reverberated conditions. For both conditions, we measure PESQ \cite{ITUT_pesq_wb_corri}, short-term objective intelligibility (STOI)\cite{taal2010short}, and the DNSMOS \cite{reddy2020dnsmos} on the enhanced speech. We furthermore measure the performance on real data using DNSMOS on real recordings of the preliminary test set $\mathcal{D}^\mathrm{real, dev}_\mathrm{DNS3}$ from DNS3 \cite{reddy2021interspeech}.

The final evaluation using a subjective listening test following the P.835 framework \cite{ITUT_P835} is carried out by the DNS Challenge organizers based on a blind test set $\mathcal{D}^\text{test}_\text{DNS3}$, which has not been seen in any training, development, or optimization process.
\vspace*{-2mm}
\subsection{Results and Discussion}
\label{sec:3.3}
\begin{table}[t]
	\large
	\centering
	\caption{\textbf{\textit{Perceptual quality measure} DNSMOS \cite{reddy2020dnsmos}} on the \textbf{\textit{current}} DNS Challenge \textbf{\textit{preliminary test set}} $\mathcal{D}^\mathrm{real, dev}_\mathrm{DNS3}$, evaluated on \textbf{\textit{real recordings}}. Best results are in {\bf bold} font, and the second best are \underline{underlined} (oracle experiment excluded).}
	\vspace*{-2mm}
	\setlength\tabcolsep{3pt}
	\resizebox{1\linewidth}{!}{
		\begin{tabular}{c c c}
			\hline
			&Methods & \stz{DNSMOS}\\ \hhline{---}
			\multirow{3}{*}{\rotatebox{90}{REF\ }}& \multicolumn{1}{l}{Noisy}            &  \stz{2.90}  \\ \hhline{~~~}
			& \multicolumn{1}{l}{\stz{DNS3 Baseline \cite{braun2020data}, trained with synthetic data only}}             &  \stz{3.24}  \\ \hhline{~~~}
			& \multicolumn{1}{l}{\stz{FCRN, with loss \eqref{MT} \cite{strake2020DNS}, trained with synthetic data only}}             &  \stz{3.30}  \\ \hhline{---}
		\multirow{4}{*}{\rotatebox{90}{New\ }}	& \multicolumn{1}{l}{\stz{FCRN/PESQNet, losses \eqref{total} and \eqref{PESQNet}}, $\langle0\!-\!2\!-\!50\rangle$ (synth.\ data only)}             &  \stz{\underline{3.31}}  \\ \hhline{~~~}
		& \multicolumn{1}{l}{\stz{FCRN/PESQNet, losses \eqref{total} and \eqref{PESQNet}, $\langle1\!-\!1\!-\!50\rangle$ , $\alpha\!=\!0.8$}}            &  \stz{\underline{3.31}}  \\ \hhline{~~~}
		& \multicolumn{1}{l}{\stz{FCRN/PESQNet, losses \eqref{total} and \eqref{PESQNet}, $\langle1\!-\!1\!-\!50\rangle$, $\alpha\!=\!0.9$}}            &  \stz{\bf 3.33}  \\ \hhline{~~~}
		& \multicolumn{1}{l}{\stz{FCRN/PESQNet, losses \eqref{total} and \eqref{PESQNet}, $\langle1\!-\!1\!-\!50\rangle$, $\alpha\!=\!0.9$ (oracle)}}            &  \stz{3.36}  \\ \hhline{---}
			
		\end{tabular}
	}
	\label{DNS3_dev}
	\vspace*{-2mm}
\end{table}

In Table\,\ref{DNS1_dev} we evaluate the performance of our novel weakly supervised training by performing an instrumental measurement on the synthetic test set $\mathcal{D}^\text{synth, dev}_\text{DNS1}$. As references, we report the performance of the DNS3 baseline \cite{braun2020data} and of the FCRN employing only the synthetic loss \eqref{MT} \cite{strake2020DNS}, both being trained with the synthetic dataset $\mathcal{D}^\text{synth}_\text{DNS3}$. Finally, the performance is measured for the FCRN trained with our novel loss $J^\text{total}_u$ \eqref{total}, employing either synthetic data only, or both synthetic and real data. Please note that the results in Table\,\ref{DNS1_dev} are measured on synthetic data $\mathcal{D}^\text{synth, dev}_\text{DNS1}$, and therefore only serve to prove that our novel proposed framework using synthetic and real data does not deteriorate under synthetic data. We observe that the DNS baseline offers the worst perceptual quality (PESQ, DNSMOS) for the conditions without reverberation. This can be attributed to an insufficient SNR improvement, reflected by an about $2$ dB lower $\Delta\text{SNR}_\text{seg}$ compared to other methods. Under reverberated conditions, the baseline offers the worst dereverberation performance reflected by the lowest SRMR score, but is good in DNSMOS. The FCRN trained with synthetic loss \eqref{MT} offers the highest perceptual quality and intelligibility reflected by PESQ and STOI scores, respectively, under both reverberation conditions. Our proposed FCRN/PESQNet trained with losses \eqref{total} and \eqref{PESQNet}, but only using synthetic data ($\langle0\!-\!2\!-\!50\rangle$ protocol) shows good balanced performance with six out of eight \nth{1} and \nth{2} ranked metrics. When including real training data as well (proposed $\langle1\!-\!1\!-\!50\rangle$ protocol, with $\alpha\!=\!0.9$), still six \nth{1} and \nth{2} ranks are obtained, which shows that including real data in training doesn't considerably harm performance on synthetic dataset $\mathcal{D}^\text{synth, dev}_\text{DNS1}$. Further decreasing $\alpha$ in \eqref{total} to $0.8$ degrades the performance to only three \nth{2} ranked metrics.

To evaluate the performance on real data, our actual goal, we measured the DNSMOS score on the real recordings from $\mathcal{D}^\mathrm{real, dev}_\mathrm{DNS3}$ as shown in Table\,\ref{DNS3_dev}. Even though only a small amount of real data (4 hours) was added to the training material, our novel loss $J^\text{total}_u$ \eqref{total} with $\alpha\!=\!0.9$ employing both real and synthetic data offers the highest DNSMOS score $(3.33)$ among all methods (except for the oracle experiment). This indicates that by training our FCRN with the help of the non-intrusive {\tt PESQNet} and the novel $J^\text{total}_u$ \eqref{total}, real data can be advantageously employed during training, which can improve the DNS performance in a real test environment. The setting with $\alpha\!=\!0.8$ shows that, for such a benefit, a high weight on the original synthetic loss $J^\text{synth}_u$ \eqref{MT} is still needed. With $\alpha=0.9$, however, {\it we are able to prove that our $\langle1\!-\!1\!-\!50\rangle$ protocol with the reference-free {\tt PESQNet} indeed solves the earlier observed issues reported by Fu et al.\ \cite{fu2019learning} with their proposed {\tt Quality-Net}. Accordingly, at this point we already can conclude that with our proposed $\langle1\!-\!1\!-\!50\rangle$ protocol and the reference-free {\tt PESQNet} indeed is able to make use of real data.} As an oracle experiment, we trained our FCRN/PESQNet using the novel training protocol $\langle1\!-\!1\!-\!50\rangle$ with $\alpha\!=\!0.9$, employing the data from $\mathcal{D}^\mathrm{real, dev}_\mathrm{DNS3}$ not only for evaluation, but also as the real data part during training. The further increased DNSMOS of $3.36$ shows that the additional information from real data can be advantageously used by employing the proposed training protocol. 

Comparing to the reference training protocol $\langle0\!-\!2\!-\!50\rangle$ employing only synthetic data, the performance improvements of our proposed training protocol $\langle1\!-\!1\!-\!50\rangle$ with $\alpha\!=\!0.9$, reflected by a $0.02$ points higher DNSMOS, are only sight. In this work, however, we are limited to a very small amount of real data (4 hours) used in training, when compared to the amount of synthetic data (100 hours). Accordingly, during training, all the real data has to be repeatedly used for several times even during one epoch of synthetic data, which may limit the performance, and especially the generalization capability, on real data. In turn, we believe that the use of much more real data will provide an even more significant performance gain.
\begin{table}[t]
	\scriptsize
	\centering
	\caption{\textbf{\textit{Subjective quality results}} in terms of \textbf{\textit{MOS scores}} according to ITU-T P.835 on the \textbf{\textit{blind test set}} $\mathcal{D}^\mathrm{test}_\mathrm{DNS3}$. Extracted from the DNS Challenge's Track 1 (real-time track for wideband signal). Best results are marked in {\bf bold}.}
	\vspace*{-2mm}
	\setlength\tabcolsep{2pt}
		\begin{tabular}{c c}
			\hline
			Submissions & \stz{Overall MOS}\\ \hhline{--}
			\multicolumn{1}{l}{\stz{Noisy}}  & 2.77   \\ \hhline{~~}
			\multicolumn{1}{l}{\stz{DNS3 Baseline \cite{braun2020data}}, trained with synthetic data only}             &  \stz{3.07}\\ \hhline{--}
			\multicolumn{1}{l}{\stz{FCRN/PESQNet, losses \eqref{total}} and \eqref{PESQNet}, $\langle1\!-\!1\!-\!50\rangle$, $\alpha\!=\!0.9$} &  \stz{\bf 3.16}\\ \hhline{--}
			
		\end{tabular}
	\label{DNS3_test}
	\vspace*{-2mm}
\end{table}

Table\,\ref{DNS3_test} presents the overall MOS scores from the subjective listening test carried out by the DNS3 organizers based on the blind test set $\mathcal{D}^\text{test}_\text{DNS3}$, which contains both synthetic and real data. Our proposed framework outperforms the DNS3 baseline \cite{braun2020data}, which is trained only with synthetic data, by $0.09$ MOS points. To some extent, this also confirms the DNSMOS results shown in Table\,\ref{DNS3_dev}, where our proposed framework also outperforms the DNS3 baseline by $0.09$ DNSMOS points. Furthermore, our proposed framework delivers a natural-sounding background noise reflected by the reported background noise MOS of $4.33$, leading to an improvement of $0.45$ MOS points over the DNS3 baseline.
\section{Conclusions}
In this paper we illustrate the benefits obtained form training a complex mask-based FCRN for speech enhancement in a ``weakly" supervised way employing both synthetic and real data. We propose a non-intrusive {\tt PESQNet}, providing a reference-free perceptual loss for the training of speech enhancement neural networks using real data. Furthermore, the {\tt PESQNet} and the FCRN are trained alternately to keep the {\tt PESQNet} up-to-date. Our proposed framework can offer good performance on synthetic data and improves over the reference methods on real data. In the subjective listening test of the Interspeech 2021 DNS Challenge, our proposed framework shows a $0.09$ points higher overall MOS score compared to the challenge baseline. Furthermore, our framework also improves over the baseline by $0.45$ background noise MOS points. {\it Even more importantly, however, we are the first to show how real noisy data can be used to train a noise suppression DNN without the use of a generative adversarial network (GAN).}
\clearpage
\newpage
\bibliographystyle{IEEEtran}
\ninept
\small
\bibliography{main}


\end{document}